## INTE 2014

# A physics exhibit to show the effect of the aerosol in the atmosphere on electromagnetic wave propagation

Dedalo Marchetti*

*Dipartimento di Matematica e fisica Università di Roma Tre, Via della Vasca Navale, 84, Roma RM 00146, Italy*
*INFN Sezione di Roma3, Italy*

**Abstract**

In this paper it is explained the construction and utility of a didactic exhibit about the effect of aerosol in atmosphere on electromagnetic wave propagation. The exhibit is composed by a lamp simulating the Sun, a Plexiglas case (the atmosphere), white or black panels (surface albedo), a combustion chamber to supply aerosol inside the case and other equipments. There are temperature and relative humidity of air sensors and 5 light sensors to measure direct and scattered light. It is possible to measure the cooling effect of aerosol inside the case and the increasing in scattered light.

## 1. Introduction and context

The influence of the aerosol in the atmosphere is very important in all climate models. It's well known that if carbon dioxide increases in the atmosphere, the temperature rises up. On the contrary the presence of aerosol in the atmosphere has a cooling effect: it increases the reflection of the Sun incident light in outer space.

There are different types of aerosol, which differ mostly in the size of the particles. The source of aerosol could be natural or artificial. An example of natural aerosol in the atmosphere is the Sahara Desert sand over the

* Corresponding author. Tel.: +39-333-2362654; fax: +39-06-5733-7102.
*E-mail address:* dmarchetti@fis.uniroma3.it  dedalo.marchetti@roma3.infn.it

Mediterranean Sea. An example of artificial aerosol is the industrial waste dispersed in the atmosphere or the solid particles ejected by the exhaust pipe of motor-vehicles.

An indirect effect on the atmosphere is cloud formation due to the aerosol, that reduces solar irradiation.

It was built a small chamber to analyze the effect of aerosol in a didactic contest (see photo in figure 1). The chamber can be monitored and controlled by the user. The exhibit allows to measure the amount of scattered light, variations in air temperature and the amount of relative humidity in the chamber. A lamp, in upper side of the chamber, simulates the irradiation of the Sun. Aerosol particles are produced in a controlled combustion chamber in which it is possible to burn little pieces of paper. The aerosol smoke is cooled before entering into the atmosphere chamber. It is possible to increase the amount of humidity in the chamber with a water heater. In the bottom of the atmosphere chamber there are 2 panels to simulate superficial albedo, one side is white like ice and land, the other side is black like sea.

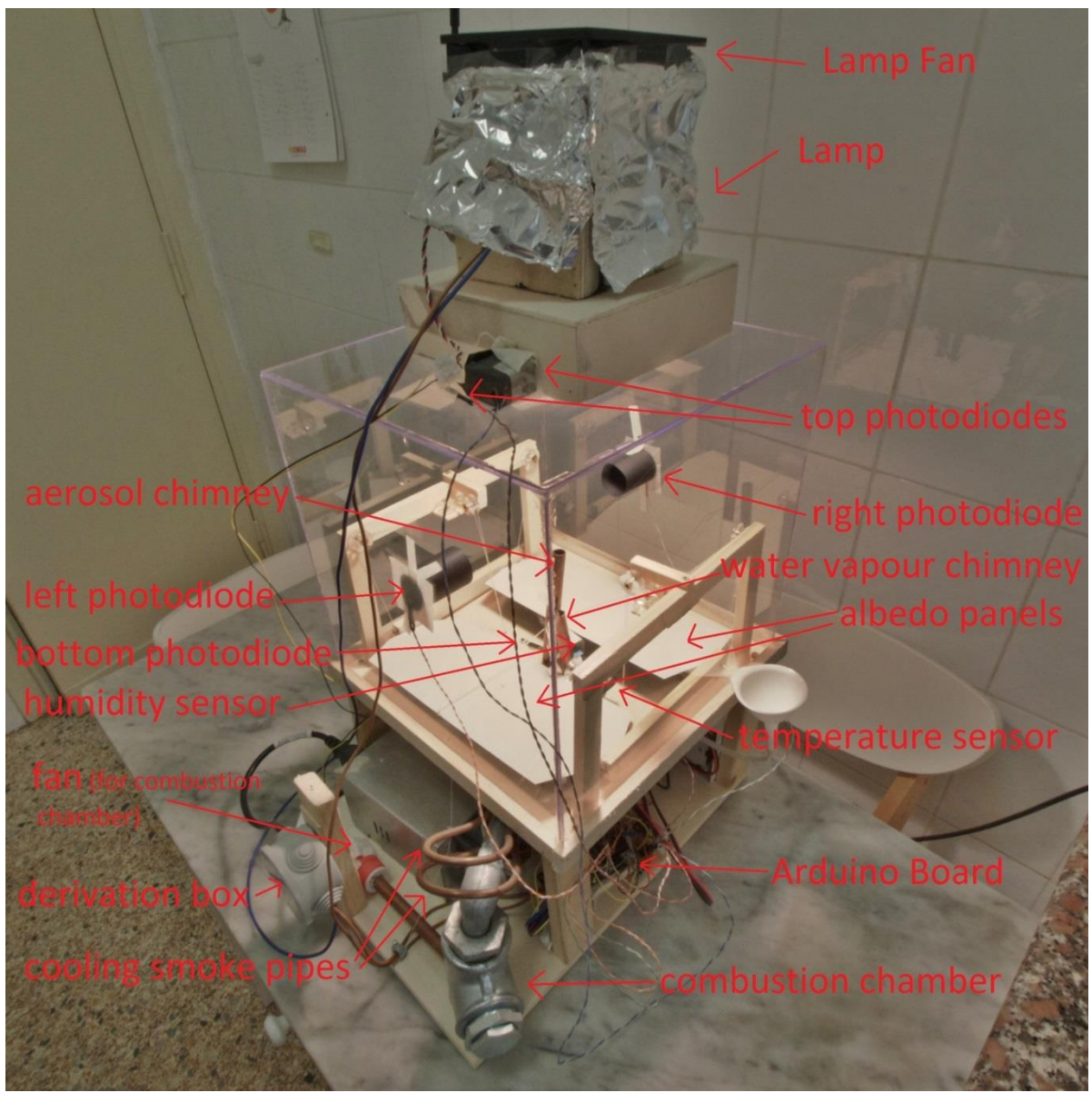

Fig. 1 - Photo of the exhibit. Red labels indicate all equipment and sensors.

## 2. General description

*2.1. Description of used equipment*

The principal component of the exhibit is a Plexiglas case (the atmosphere chamber). The shape of the case is a cube with 30 cm edge length. The case is putted on a bottom plane of poplar plywood. The air in the case represents the Earth atmosphere.

In the upper side of the case there is a common lamp with E27 connection. A 12 cm fan cools air near the lamp to protect Plexiglas case by overheating. The lamp chosen is a "spot lamp". This type of lamp has a particular coating inside the bulb to concentrate illumination in the lower part. The spectrum of artificial light is slightly different from Sun Spectrum that is similar to a black body at the temperature of 5900 K. The temperature of the filament of the lamp is between 1500 K and 2700 K, so the spectrum is similar to Sun with more light in higher wavelength i.e. infrared and minor in shorter wavelength i.e. ultraviolet. The illumination is more similar to an exoplanet that orbits around a red dwarf star.

In the bottom part of the case there are 2 panels to adjust surface albedo (see figure 2). Each panel is made gluing a white paperboard with a black paperboard together. In the middle of them there is also a long skewer between the two paperboards to rotate each panel. The white side of the panels simulates the ice on the Earth surface, on the contrary the black side simulates the sea. To move up and down and to rotate the panels, there are 4 ropes with some pulleys.

To uplift the panels: 2 ropes are rolled up around a "reel of thread" by an electric motor and a gearbox reduction placed just below the floor of the plexiglass case (see photo in figure 3).

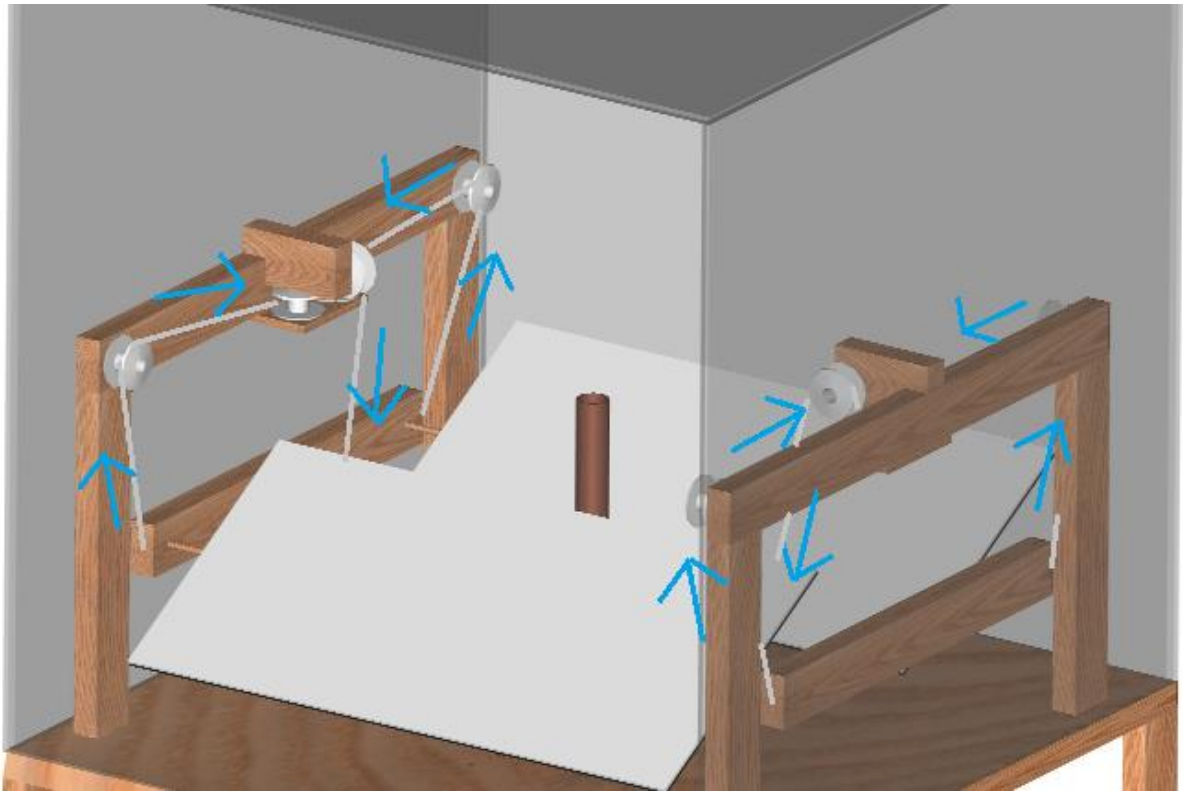

Fig. 2 – A detail of the 3D project for the moving system of albedo panels. The arrows indicate the moving direction of ropes to uplift panels.

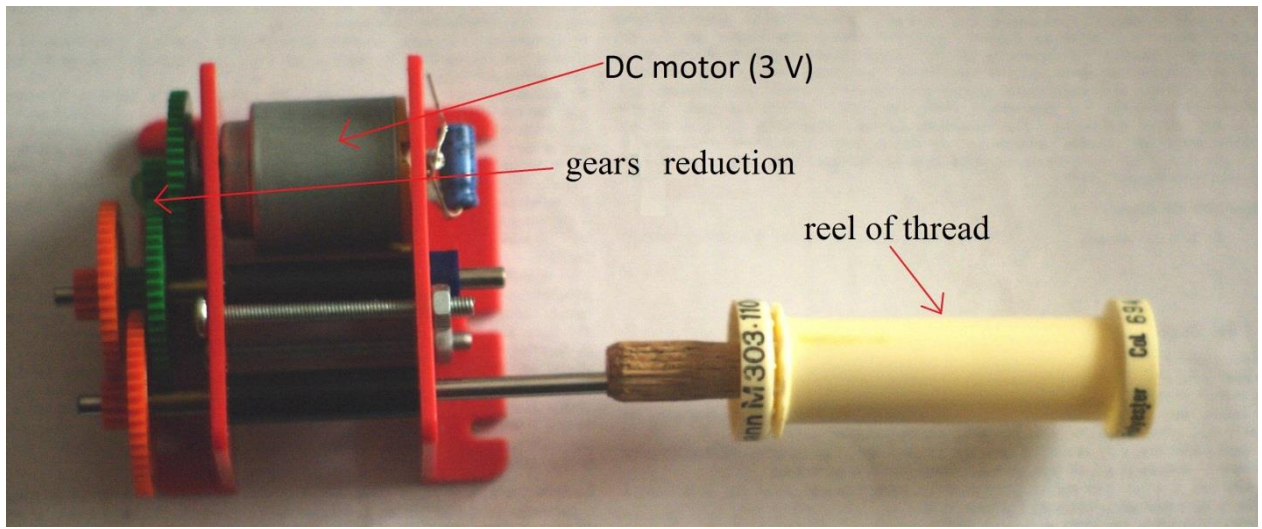

Fig. 3 - Photo of the motor and gearbox reduction of "albedo lift", before the installation into the exhibit. The "reel of thread" to roll the ropes is visible in the photo.

This motor is controlled by the integrated circuit NJM2675D, that is a power amplifier H bridge allowing the motor to run in each direction (see scheme in figure 4).

The lift system is commanded by a computer. This computer is connected with the development board Arduino Duemilanove with serial communication (USB with a virtual serial port). Arduino generates the correct sequence to uplift the panels; then it waits 15 seconds in which panels must be rotated manually with 2 another ropes; then the Arduino generates the different sequence to get down the lift system.

To produce aerosol, there is a combustion chamber below the floor of the plexiglass case. The horizontal position of the combustion chamber is shifted out from plexiglass case to avoid dangerous risks of flames or overheating. The combustion chamber is made by using a hydraulic tee threaded adaptor. The bottom connector is closed by a nut, the horizontal connection is utilized to fill fuel (it could be closed screwing a nut after the paper burn). The top connection is connected with 2 straight copper pipes to the inner part of the Plexiglas case, which are the chimney of the combustion chamber. The adaptor has been drilled down left to connect a pipe of 1 cm of diameter to pump air into the combustion chamber. The external end of the pipe is connected with a 2,5 cm x 2,5 cm fan.

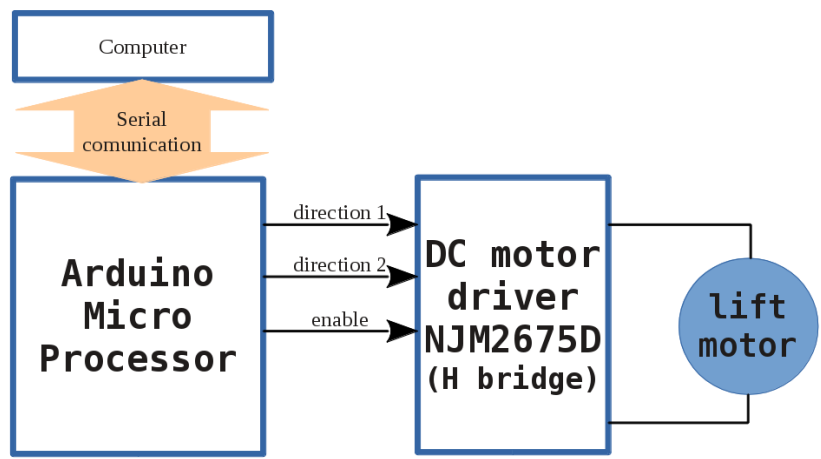

Fig. 4 - Scheme of the motor control system.

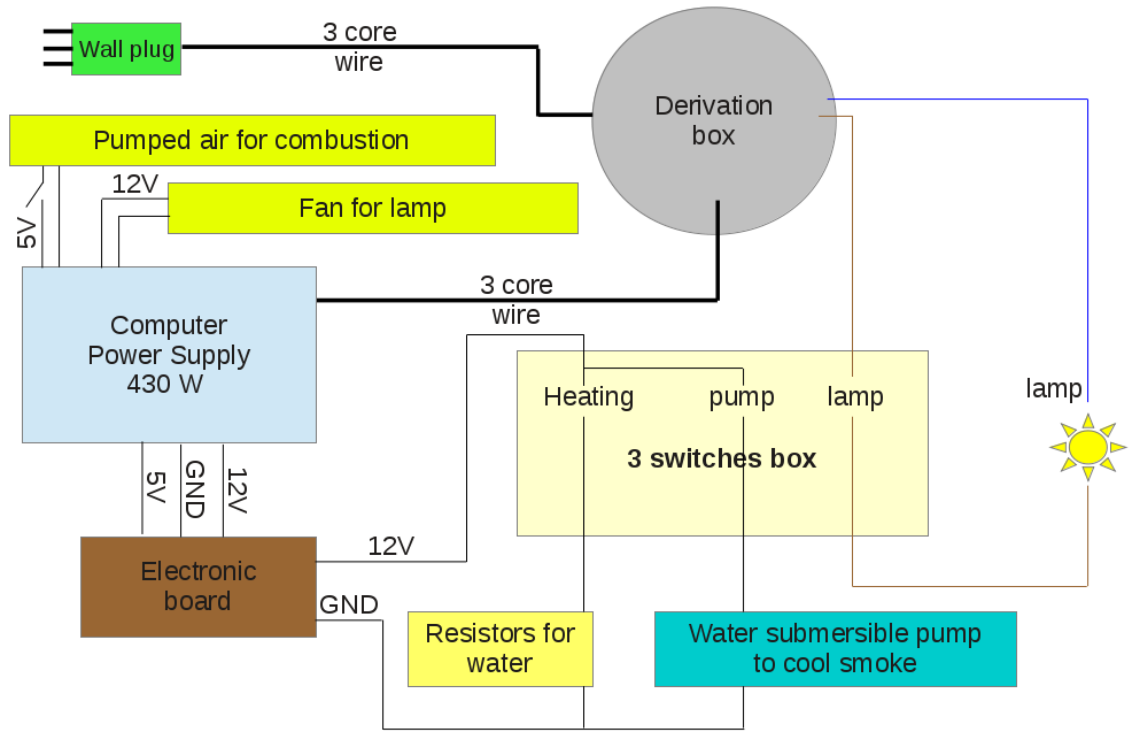

Fig. 5 - Electrical scheme of all equipment.

Little pieces of paper are used like fuel. A match is used to ignite the fire. Switching off the fan, it is possible to stop the combustion. The aerosol obtained by this combustion is similar to that one produced in a forest fire.

Around the horizontal pipe of the chimney there are 4 flexible copper pipes to cool the smoke. In these pipes flows pumped cool water, taken from a glass jar provided with a submersible pump. The pump is powered with 12 V electric tension. The jar is filled with a mix of water and ice. The pump gets water from its lower part, so the ice will not be sucked.

The cooling of the smoke is important to not change the temperature in the Plexiglas case. The pipes were chosen of copper, as this material has a high thermic transfer coefficient.

Below the floor of the case there is another glass jar to boil water for increasing water vapor in the atmosphere chamber. In the jar there are 2 electric resistors to evaporate water. The resistors are 1,8 Ohm and 2,2 Ohm resistance and power of 25 Watt. They are connected in series to a power supply of 12V. The water vapor is introduced in the upper chamber by a copper pipe connected with the lid of the jar.

An electrical scheme of all equipment is visible in figure 5. To power all equipment (except lamp) is used a "computer ATX power supply" of 350 Watt. A lower power supply is enough for this exhibit, but nowadays the power supply used is the minimum for a typical computer. The power supply is switched on, connecting the green cable with a black cable on the 24 pin Molex connector, which usually goes to the motherboard. This connection simulates the Power On Command from the motherboard to the power supply. The lamp is powered directly by 220 V AC tension and is turned ON and OFF by a switch.

A fuse is put inside a derivation box to protect all electric components from eventually short-circuits. A 3core wire with an electrical plug 220 V (starting from the derivation box) powers all the exhibit.

*2.2. Sensors to monitor the atmosphere in the chamber*

To monitor the propagation of the electromagnetic waves into the atmosphere chamber, it is possible to use photodetectors.

In this exhibit, 5 photodiodes SFH206K have been chosen to detect the intensity of light. The model of these photodiodes has an active area of 7mm$^2$ (2,65 mm x 2,65 mm), a half intensity response angle of 60° and a spectral range of sensitivity visible and near infrared.

2 photodiodes are put at top of the atmosphere chamber to measure intensity of reflected light upward. One of them has a UV/IR cut filter that passes visible light from 420 nm to 680 nm. The filter and its photodiode are putted into a black paperboard box to stop exterior light. The other top photodiode detects light in the full band of photodetector (from 400 nm to 1100 nm). It is attached directly at the external face of the Plexiglas with adhesive tape. That is possible as for this particular photodiode the sensitive face is flat.

They simulate 2 sensors putted on a satellite looking down the Earth atmosphere. 2 photodiodes are attached, in the same way, on the left and right side of the atmosphere chamber to measure the light scattered by 90° relative to the line of propagation. Due to the large angle of view of the photodiode it is necessary to put a black paperboard tube with a diameter of 3cm and a length of 5 cm to reduce acceptance angle.

One photodiode is attached with hot glue on the bottom of the chamber to see direct light. It simulates a sensor putted on the Earth surface.

A sensor LM35 is used to measure the temperature of the atmosphere chamber. It is putted on the bottom of the atmosphere chamber. The integrated circuit LM35 has an output voltage linearly proportional to the temperature. The proportional constant is typically 10 mV every Celsius degree. LM35 does not require a calibration. It is calibrated in production process at wafer level. This sensor can be powered by + 4 V to 30 V.

A digital sensor RHT01 is used to measure the relative humidity of the atmosphere in the chamber. A second measure of the temperature of the atmosphere in the chamber can be obtained by the same sensor (RHT01).

RHT01 sensor operates in a range from 20% to 90 % relative humidity and a range from 0° to 50° Celsius temperature. Its accuracy is 5% RH for humidity and 2°C for temperature. Its power's voltage should be from 3.3V to 6V. The RHT01sends data by a serial communication with "MaxDetect 1-wire" protocol.

In this exhibit a 5V tension is used to power the LM35 and the RHT01 sensors.

*2.3. Data Acquisition System*

All measures are stored in a PC with a simple data acquisition system. Data transfer is made in 2 steps. In the first step measures are acquired in the Arduino microcontroller board. In the second step data are sent to the computer, to be recorded in a file of a storage disk. The 5 photodiodes and analog output of LM35 IC are connected directly to analog pins of Arduino board. The microcontroller ATMEGA328, in the Arduino Duemilanove, has a 10-bit successive approximation analog-to-digital converter. The analog-to-digital converter is connected to a 6-channel analog multiplexer, which allows 6 analog inputs. The AREF (analog reference) pin of Arduino allows to change the maximum tension of the analog-to-digital converter: it is possible to use a voltage less than 5V to obtain higher sensitivity conversions. In this exhibit a 2 resistances divider is used to generate the analog reference voltage for Analog-to-digital converter. The resistances are 1 KOhm and 8.2 KOhm and the reference voltage is 0,54 V with 5,00 V input tension.

The digital pin of the RHT01 sensor is connected to the digital pin 4 of the Arduino board (in the Arduino setup, this pin is configured as a digital input). The values of RHT01 are read by the library released from the producer.

The same Arduino board controls the motor of the moving system of albedo panels. During this process, the acquisition of measures is pausing.

The Arduino programm written for the exhibit reads a character from serial communication with PC. This character is the command to move "albedo lift" ("l") or to read sensors ("o"). The computer waits for 2 seconds. If the user has typed a character on to the keyboard in this time, the program sends a command "l" to Arduino, else it sends the command "o". Data received from Arduino are recorded by the computer in a file. This file is plotted every 2 seconds by Gnuplot program with a script.

## 3. A run of measures taken by the exhibit

One or two hours before using the exhibit, it is necessary to switch on the lamp, which warms the air in the atmosphere chamber.

In this run, water vapor is not added in the chamber. Some ice cubes are putted in the pump water glass. The acquisition of data taken in 3 hours, 8 minutes and 19 seconds are reported in the graph in figure 6. A spreadsheet program can be used to analyze data. The calculated values are reported in table 1. All the values are the mean of time series in a specific range. The errors of the values are the standard deviations of the same series. If the standard deviation is smaller than the instrumental accuracy, the accuracy is chosen. All values are calculated with 2 different albedos (black and white), both of them are measured without aerosol and with aerosol.

Table 1. The mean values and standard deviations of the measures obtained by the exhibit. The error of temperature is not the standard deviation of data set which is minor of the accuracy.

| Sensor | No aerosol black albedo | No aerosol white albedo | With Aerosol black albedo | With Aerosol white albedo |
|---|---|---|---|---|
| Bottom photodiode [ADC counts] | 785,1 ± 6,4 | 777,5 ± 6,2 | 755,5 ± 7,7 | 756,0 ± 6,0 |
| Right photodiode [ADC counts] | 561,2 ± 3,1 | 571,5 ± 3,3 | 604,2 ± 4,4 | 609,7 ± 4,0 |
| Left photodiode [ADC counts] | 579,4 ± 3,2 | 582,6 ± 3,4 | 630,4 ± 4,3 | 632,8 ± 4,3 |
| Top photodiode [ADC counts] | 656,4 ± 5,4 | 691,9 ± 5,2 | 686,4 ± 6,9 | 698,1 ± 5,2 |
| Top photodiode visible range [ADC counts] | 546,0 ± 1,8 | 562,3 ± 2,6 | 555,8 ± 4,2 | 562,9 ± 2,3 |
| Temperature LM35 [°C] | 28,6 ± 0,5 | 28,0 ± 0,5 | 26,4 ± 0,5 | 26,2 ± 0,5 |

The white-albedo measure of the light, acquired by the bottom photodiode is compatible with the black-albedo measure. These two measures could be compatible why this sensor is beside of the albedo panels.

The measures of the light, acquired by the right and left photodiodes increase when the aerosol is introduced into the atmosphere chamber. In fact the aerosol particles interact with light, scattering towards these photodiodes.

The white-albedo measures of the light, acquired by the top photodiodes have a higher intensity than the black-albedo measures.

When the aerosol is introduced into the atmosphere chamber, the top photodiodes measure an increment of light intensity, the bottom photodiode measures a decrease of light intensity and the LM35 thermometer measures a decrease of temperature.

These results may explain the aerosol effect in the Earth atmosphere: the reduction of Earth surface intensity of Sun light, and a lower temperature of the atmosphere, in opposition to the greenhouse effect.

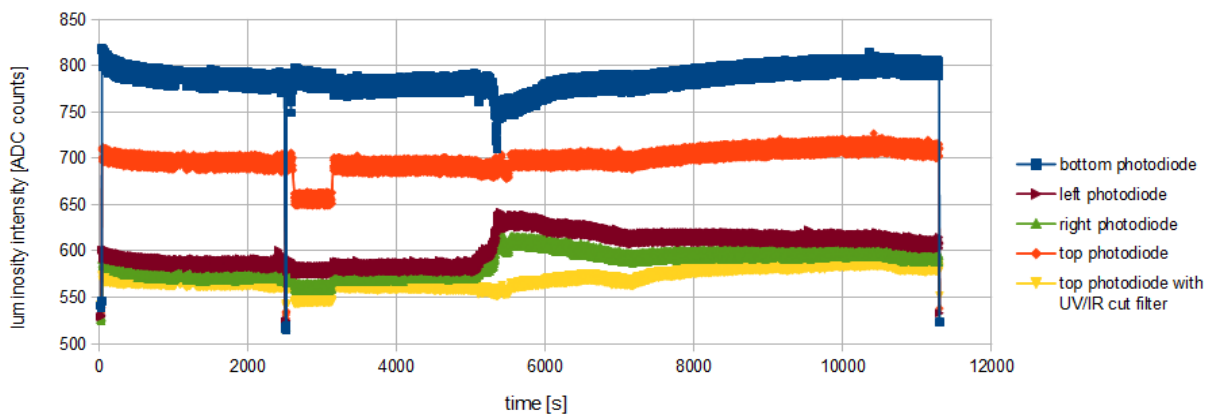

Fig. 6 - Graph of the time series of bottom, left, right and top photodiodes. The 3 short low peaks are due to switch off the lamp for a few seconds.

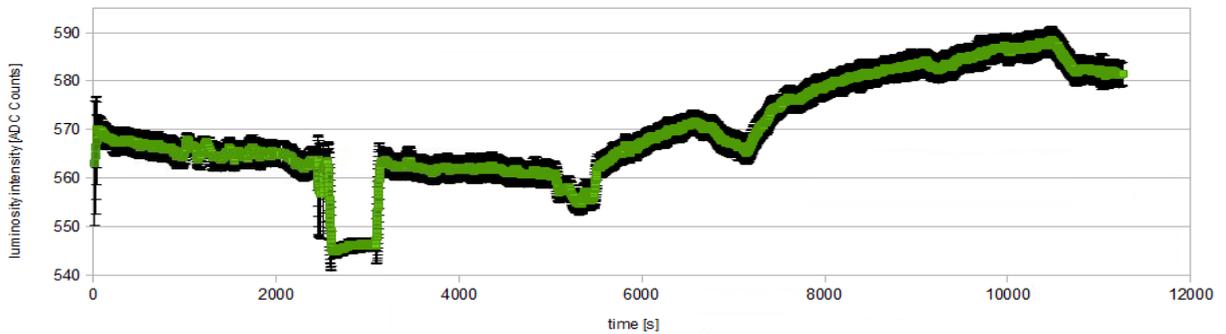

Fig. 7 - The graph of data acquired by the top visible photodiode with "mean filter"

A mean filter can be used to display data emphasizing their trend. The data of the top visible light photodiode are processed with a 20 points "mean filter" (see graph in figure 7).

The filter calculates the mean of 20 data as in equation (1) for all data except last 19 data of the time series:

$$y_n = \frac{\sum_{k=1}^{10} x_{n+k}}{20} \tag{1}$$

From 2624s to 3150s and from 5435s to 5515s the Albedo panels are black; in the other intervals of time they are white. The aerosol smoke is introduced in the atmosphere chamber from 5074s. A photo of aerosol in the atmosphere chamber is visible in figure 8.

24 s and

## 4. Conclusion

The exhibit is useful for a didactic purpose. It is also possible to see special trajectories of the aerosol smoke. The temperature and humidity sensors can be used for thermodynamic experiments.

It could be either a good incentive to study physics in a secondary school or a representation of some physics principles in a scientific museum.

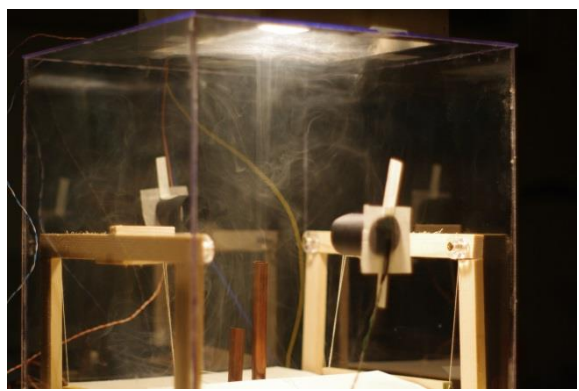

Fig. 8 - A photo of the aerosol when is introduced in the chamber.